\documentstyle[aps,prl,twocolumn]{revtex}
\begin{document}
\twocolumn[\hsize\textwidth\columnwidth\hsize\csname@twocolumnfalse\endcsname]

{\bf Zaikin et al. reply:}
In a recent Letter~\cite{ZGOZ} we reported on a microscopic theory
of quantum phase slips (QPS) in ultrathin superconducting wires
and arrived at conclusions that are {\it qualitatively} different
from those previously reached in Ref.~\cite{Duan}.
The Comment~\cite{DuanC} in turn disagrees with our conclusions.

For clarity, we summarize the main differences between the
work of Ref.~\cite{Duan} and ours~\cite{ZGOZ}. 
1) We find a finite QPS fugacity, whereas Ref.~\cite{Duan} finds a
vanishingly small value $\sim\exp(-137)$. 2) In the physical limit,
we obtain a $log$ -- interaction between QPS's, related to the presence
of the (acoustic in 1D) Mooij-Sch\"{o}n mode~\cite{ms}. In contrast,
Ref.~\cite{Duan} considers the ``cosmic string'' limit with the
``log(log)'' interaction \cite{Zhang}. 3) In contrast to \cite{Duan},
we find a new superconductor to metal phase transition in 1D superconducting
wires, due to proliferation of quantum phase slips.
In reaction to the points raised in Ref.~\cite{DuanC}:

i) Ref.\cite{DuanC} argues that our theory~\cite{ZGOZ} is phenomenological
because ``Once an order parameter was assumed, the model immediately ceases
to be a microscopic one''. We note that ``real time'' non-equilibrium
superconductivity, the {\it microscopic} theory for far-from-equilibrium
states and dynamics at the time scale of the inverse gap, has been firmly
established and {\it is} formulated in terms of the order parameter~\cite{LO}.
Furthermore, in our work~\cite{ZGOZ} we {\it do not assume} any order
parameter but rather {\it derive} the effective action as a functional
of the order parameter field (see also~\cite{ogzb} for more details).
Our theory describes quantum tunneling processes which involve electronic
states far from, not ``near'' equlibrium.
Ref.\cite{Duan} considers only the geometric quantities $L,C$ and ignores
the kinetic ones $\tilde{L},\tilde{C}$ whereas we include all of them.
For typical system parameters we always have $\tilde L \gg L$. This is 
the key reason for the electromagnetic contribution to be by a factor 
$\sim \sqrt{S}/\lambda_L$ smaller than that obtained in \cite{Duan} 
($S$ is the wire cross section and $\lambda_L$ is the London length). 

ii) Interactions between QPS's depend on the wire parameters. For the thin 
wires with $\sqrt{S} < \lambda_L$ considered in~\cite{ZGOZ}
we discern three regimes: 1) Finite length wires with constant geometric
capacitance C. The inter-QPS interaction is purely logarithmic \cite{ZGOZ}. 
2) Finite length wires with $C(k)\propto 1/\ln(1/k)$.
The interaction is proportional to $\sqrt{\ln(1/k)}$.
3) Astronomically long wires of a size $X \gg x_0 \exp (2\pi \lambda_L^2/S)$. 
Depending on $C$ the interaction is of a ``$\sqrt{\ln }$'' or of 
a ``log(log)'' form.

The limit 3) as considered in Refs.~\cite{Duan,DuanC} is irrelevant:
for $\sqrt{S} \sim \lambda_{L}$ it
can be realized only for wires of length $\gg 1$ cm (the longest wires
in Refs.~\cite{Gio,Dynes} did not exceed $10^{-2}$ cm).
For $\sqrt{S} \sim 10$ nm the length, where a log(log) behavior shows up
is $X \sim x_0 \exp (600)$. This is too big even for cosmic strings. 

The limit 2) (overlooked in Refs.~\cite{Duan,DuanC}) -- although
theoretically possible -- is of limited experimental relevance as well.
The important point is that in practice stray capacitances cut off the
$k$-dependence in $C(k)\propto 1/\ln(1/k)\to 1/\ln(d)$ at a scale $d$ that
depends on the experimental details. This may influence only the
short-distance behavior.

iii) Our results for both the core and the electromagnetic parts of
the QPS action~\cite{ZGOZ} are {\it parametrically} different from 
those of~\cite{Duan}. Our core energy essentially depends on the
wire normal state conductivity (which characterizes dissipation) and
the velocity of the mode~\cite{ms}. The result~\cite{Duan} 
does not contain these parameters.

In Ref.~\cite{Duan} the electromagnetic contribution to the QPS action 
was found to be of order of the inverse fine structure constant 
$\sim 1/\alpha\sim 137$. This result is counterintuitive, as it is 
independent of the wire thickness: e.g. it remains constant
for vanishing thickness $S \to 0$, an obvious impossibility.
As explained under i), including the kinetic inductance  $\tilde{L}$
{\it reduces} the EM barrier of \cite{Duan} by a factor $\sqrt{S}/\lambda_L$.
For wires with thickness in the 10 nm range \cite{Gio,Dynes} 
this gives a QPS fugacity $\sim \exp (- 10)$, making QPS phenomena 
observable reality.

iv) We do not understand why~\cite{DuanC} prefers to quote only the earlier
data by Giordano on thicker wires, in~\cite{Gio} results for wires with
radii down to $\simeq 8$ nm are reported.
In~\cite{ZGOZ}
we point out that for the thinner wires~\cite{Gio} our theory yields results
{\it consistent} with experimental findings~\cite{Gio}. These wires are 
reported to be homogeneous at least on scales $\gtrsim$ 10 nm~\cite{Gio}.
Also, Ref.~\cite{Dynes} reports deviations from the
thermally activated phase-slip predictions: Fig. 5 of~\cite{Dynes}
shows that (we quote) ``The LAMH fits systematically deviate with
decreasing thickness and wire width.''.
Furthermore, they do observe a {\it finite} zero temperature resistivity:
``If we extrapolate our data in the 220 \AA~ wire in Fig.2 to $T=0$, then
for thicknesses that are just on the superconducting side of the
transition, we will obtain a finite value of $R$.''. In this sense 
the observations~\cite{Dynes} and~\cite{Gio} are qualitatively similar.
   
A.D. Zaikin$^{1,2}$, D.S. Golubev$^{2}$, 
A. van Otterlo$^{3}$, and G.T. Zim\'{a}nyi$^{3}$

$^1$ Institut f\"{u}r Theoretische Festk\"orperphysik, Universit\"at
Karlsruhe, 76128 Karlsruhe, FRG, $^2$ P.N.Lebedev Physics Institute,
117924 Moscow, Russia, $^3$ Physics Dept.,
University of California, Davis, CA 95616, USA

\end{document}